\documentclass[conference]{IEEEtran}
\IEEEoverridecommandlockouts
\usepackage{amsmath,amssymb,amsfonts}
\usepackage{algorithmic}
\usepackage{graphicx}
\usepackage{textcomp}
\usepackage{xcolor}
\usepackage[hidelinks]{hyperref}
\usepackage[style=ieee, maxbibnames=6]{biblatex}
\usepackage{url}

\usepackage{float} 
\usepackage{subcaption} 
\usepackage{listings}

\def\BibTeX{{\rm B\kern-.05em{\sc i\kern-.025em b}\kern-.08em
    T\kern-.1667em\lower.7ex\hbox{E}\kern-.125emX}}
\begin{document}

\title{Optimizing STAR Aligner for High Throughput Computing in the Cloud}

\author{\IEEEauthorblockN
{
Piotr Kica\IEEEauthorrefmark{1}\IEEEauthorrefmark{2},
Sabina Lichołai\IEEEauthorrefmark{1}\IEEEauthorrefmark{3},
Michał Orzechowski\IEEEauthorrefmark{1}\IEEEauthorrefmark{2}\IEEEauthorrefmark{3},
Maciej Malawski\IEEEauthorrefmark{1}\IEEEauthorrefmark{2}
}
\IEEEauthorblockA{\IEEEauthorrefmark{1}Sano Centre for Computational Medicine, Krak\'ow, Poland (\url{https://sano.science/})\\}
\IEEEauthorblockA{\IEEEauthorrefmark{2}Faculty of Computer Science, AGH University of Krak\'ow, Poland\\}
\IEEEauthorblockA{\IEEEauthorrefmark{3}ACC Cyfronet, AGH University of Krak\'ow, Poland\\}
}

\maketitle

\begin{abstract}

We propose a scalable, cloud-native architecture designed for Transcriptomics Atlas Pipeline, using a resource-intensive STAR aligner and processing tens or hundreds of terabytes of RNA-seq data. We implement the pipeline using AWS cloud services, introduce performance optimizations and perform experimental evaluation in the cloud. Our optimization techniques result in computational savings thanks to  the "early stopping" approach, selection of right-sized resources, and using newer version of Ensembl genome.

\end{abstract}

\section{Introduction}

Transcriptomics Atlas pipeline is a data- and compute-intensive pipeline, based on a sequence aligner -- STAR~\cite{dobin2013star} -- that processes tens or hundreds of terabytes of RNA-seq data. STAR is a very accurate and also resource intensive tool, which can benefit from using cloud resources, but requires a well designed architecture and several performance optimizations to maximize its efficiency. In this paper, our goals are:
\begin{itemize}
    \item Development of a cloud-native solution for running STAR aligner in the cloud, exploiting scalability, cost-efficiency and on-demand resource availability,
    \item High utilization of resources and pipeline efficiency,
    \item Minimization of cloud costs.
\end{itemize}

\section{Pipeline and cloud architecture}
\label{sec:arch}

\begin{figure}[b]
    \centering
    \includegraphics[trim={0.7cm 0 0.7cm 0},clip,scale=0.43]{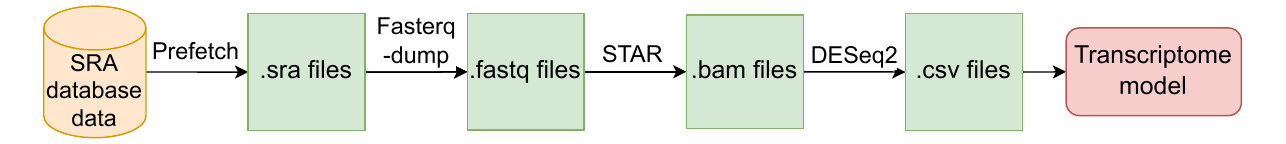}
    \caption{Transcriptomics Atlas Pipeline for STAR.}
    \label{fig:TAtlas_pipeline_STAR}
\end{figure}

The Transcriptomics Atlas pipeline consists of four steps:
\begin{enumerate}
    \item Downloading \textit{SRA} file using \textit{prefetch} tool.
    \item Converting into \textit{FASTQ} file using \textit{fasterq-dump} tool.
    \item Alignment of reads using \textit{STAR}.
    \item Count normalization using \textit{DESeq2}.
\end{enumerate}
The general pipeline is presented in Fig.~\ref{fig:TAtlas_pipeline_STAR}. We are running STAR version 2.7.10b with \mbox{"--quantMode~GeneCounts"} option. STAR aligner gives us highly reliable results and allows extensive customization of alignment parameters. Current implementation of the Transcriptomics Atlas project is publicly available on GitHub under the MIT license~\cite{neardata_repo}. 

STAR~\cite{dobin2013star} is a software that allows alignment of large transcriptome RNA-sequences. It is a well-established and accurate aligner in bioinformatics. Usually it requires a high amount of RAM to run (tens of GiBs, depending on the size of a reference genome). This aligner, as many others, requires a precomputed genomic index data structure which in this case is fully loaded into memory.

Data is obtained from the NCBI SRA repository~\cite{NCBI_SRA}, which contains more than 30PB of sequencing data. We select nucleotide sequence data generated by human samples, on the basis of tissue/cell and appropriate technical parameters. We aim to process the subset consisting of at least 7216 files and 17TB of \textit{SRA} data. 

\label{sec:input_dataset}

\begin{figure}[b]
    \centering
    \includegraphics[scale=0.4]{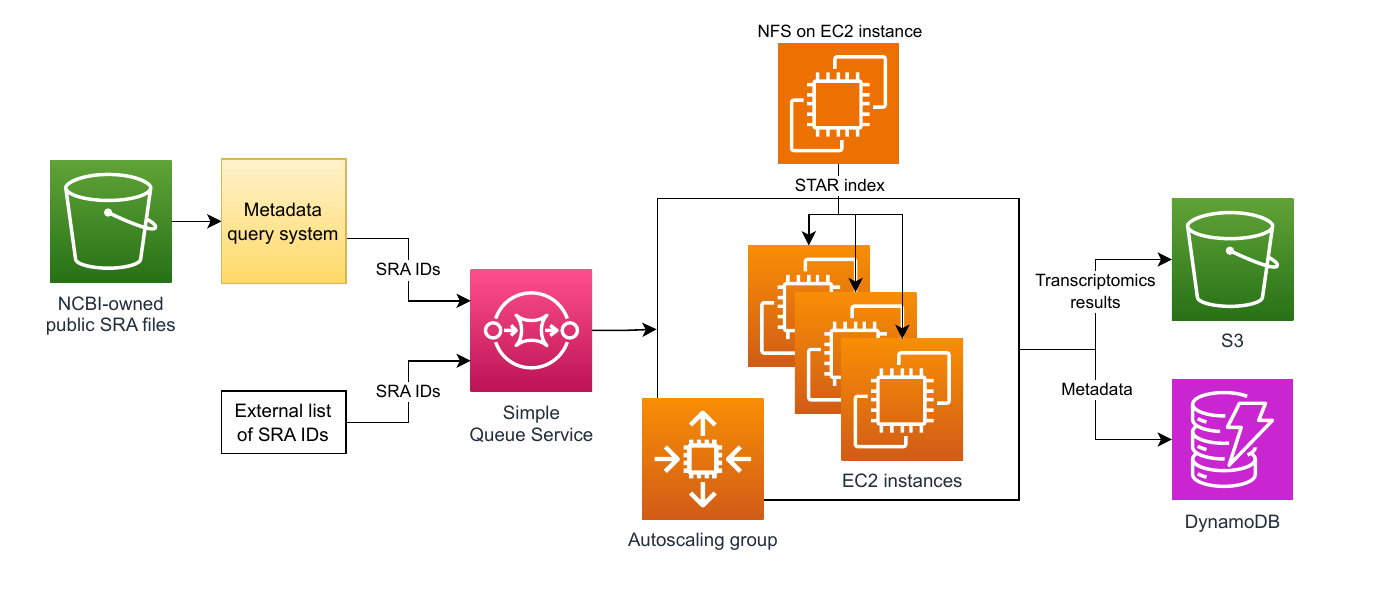}
    \caption{Cloud architecture for Transcriptomics Atlas Pipeline.}
    \label{fig:cloud_architecture}
\end{figure}

In order to create a cloud-native and efficient architecture we first gathered pipeline requirements, performed small scale experiments, and analyzed the limits of cloud services. The designed AWS cloud architecture for the Transcriptomics Atlas pipeline is presented in Fig.~\ref{fig:cloud_architecture}. We decided to use EC2 instances as the main compute resource to create a dynamic virtual cluster. Those instances are scaled using AutoScalingGroup and can be run in spot mode for cheaper processing. Each instance is launched from a predefined configuration and image that contains the software required for the pipeline. Each instance downloads the pre-computed STAR index and loads it into system memory during the initialization phase. The SRA IDs that require processing are sent into an SQS queue from which instances poll. When the pipeline has completed successfully, the pipeline results are uploaded to an S3 bucket. 

\section{Application-specific Optimizations}
\label{sec:app-opt}

\subsection{Ensembl Genome: Release 108 versus Release 111}

In order to create a Transcriptomics Atlas we are required to use the Ensembl "toplevel" genome type as it contains all known contigs, both aligned to chromosomes and any non-localized or unlocated scaffolds, which are not present in an alternative "primary\_assembly" genome type.
Between releases, especially 109 and 110, a significant number of unlocalized sequences were assigned to specific sites on the chromosome, which significantly reduced the size of the \textit{FASTA} genomic files, as well as simplifying the index. 

\begin{figure}[t]
    \centering
    \includegraphics[scale=0.355]{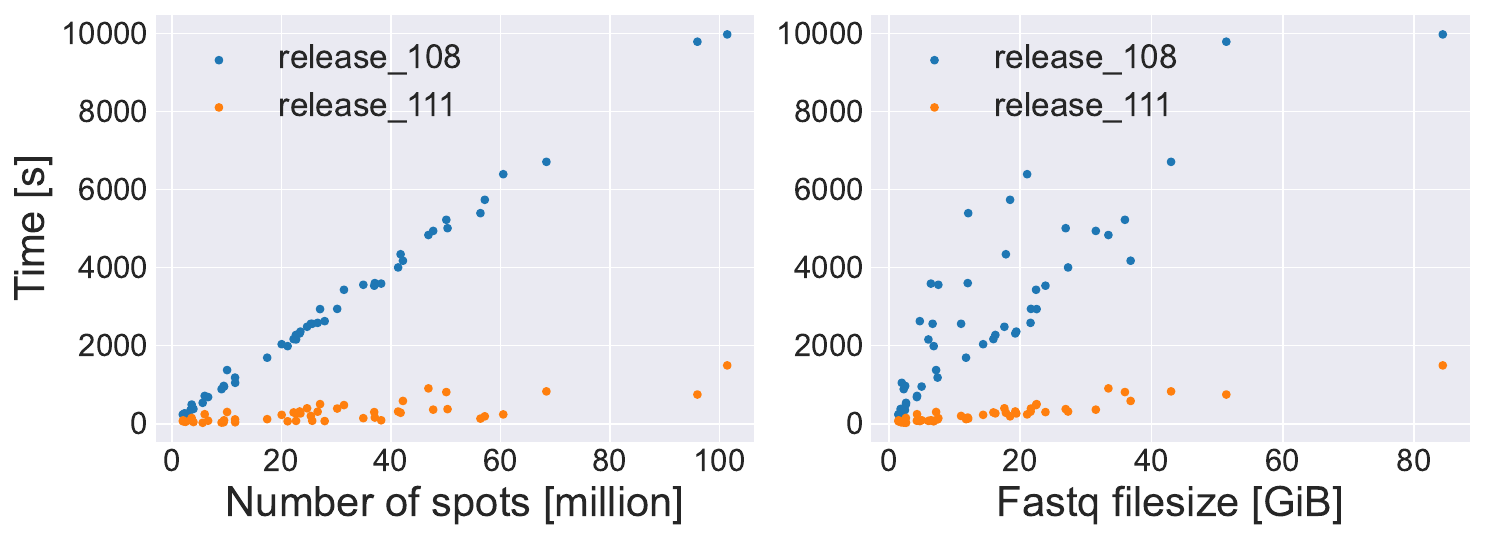}
    \caption{STAR execution time with index generated on different genome releases.}
    \label{fig:STAR_release_times}
\end{figure}

We performed an experiment on a single VM using genomes generated on releases 108 and 111 using "toplevel" sequences. 
The results show that the index generated on release 111~\cite{ensembl_release111} of the human genome gives a great improvement in execution time - more than \textbf{12 times faster} on average (weighted by \textit{FASTQ} size). The  values are presented in Fig.~\ref{fig:STAR_release_times}. Also the newer version is significantly smaller (29.5GiB vs 85GiB) which decreases computational requirements. This optimization will significantly reduce the cost of running the pipeline.

Test configuration:
\begin{itemize}
    \item Instance Type: r6a.4xlarge (16 vCPU, 128GB RAM)
    \item Input: 49 \textit{FASTQ} files (15.9 GiB mean size, 777GiB total)
    \item Index size: 85 GiB (release 108), 29.5 GiB (release 111)
\end{itemize}

Besides the speedup, using a much smaller index allows us to use smaller and cheaper instances, reduces the initial overhead associated with downloading and loading index to shared memory - all this acquiring nearly identical mapping rate ($<$1\% mean difference). Version 110 has been released on 04.2023 and not everyone may have migrated to a newer version which is recommended due to computational savings.

\subsection{Early stopping for STAR alignment}

\begin{figure}[t]
    \centering
    \includegraphics[scale=0.275]{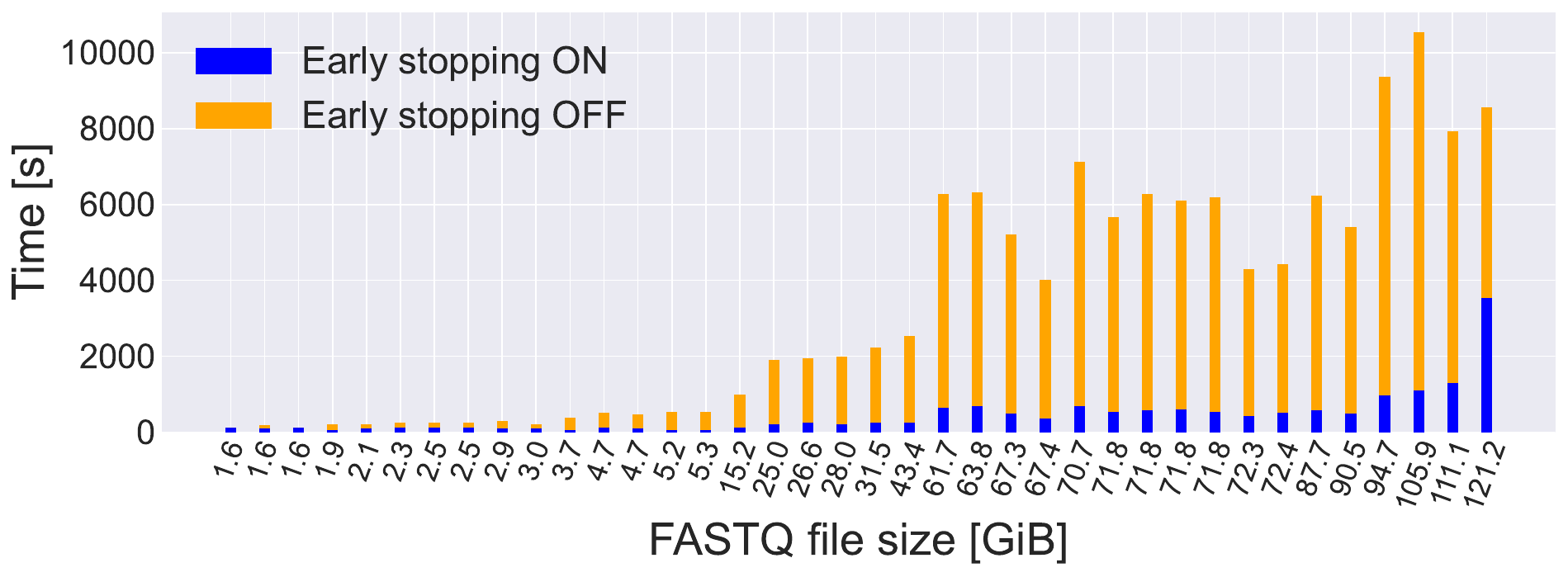}
    \caption{Time savings due to early stopping feature. Yellow bar represents unnecessary compute time.}
    \label{fig:early_stopping}
\end{figure}

STAR aligner produces a \textit{Log.progress.out} file which reports job progress statistics including current percent of mapped reads. Our goal is to create a comprehensive dataset of processed files for which the STAR aligner returns an acceptable mapping rate (above 30\%). By analyzing 1000 of \textit{Log.progress.out} files we identified that processing at least 10\% of the total number of reads is enough to decide whether the alignment should be continued or it should be aborted due to insufficient mapping rate. The analysis showed that with such approach we can safely early terminate 38 (out of 1000) alignments. Terminating early these STAR alignments would result in about \textbf{19.5\% reduction in total STAR execution time} (30.4h out of 155.8h). In Fig.~\ref{fig:early_stopping} we see when an early stopping termination would take place. The inputs identified for termination turned out to be single cell sequencing data, which seems to be a suboptimal option for our task due to the lack of complete mRNA coverage within the library. Therefore, this filtering step allows us to exclude this type of data early in the pipeline. With this approach we reduce wastage of  resources and increase overall throughput. 

\section{Conclusions and Future Work}
\label{sec:conclusions}

This work presented our cloud architecture for high-throughput computing with the STAR aligner as its core. We observed that using a newer genome release of the "toplevel" type greatly reduces pipeline requirements and execution time. Early stopping optimization we proposed notably increases the pipeline throughput, which suggests that other (pseudo)aligners should also provide the current mapping rate value (e.g. Salmon does not). Those insights are applicable outside the cloud environment (HPC or workstations). Faster STAR processing time will reduce time to make a diagnosis time by a geneticist which runs similar STAR alignment on patient's RNA-seq data. Further research will measure applicability of those findings for other aligners.

\section*{Acknowledgment}
The publication is supported by the Polish Minister of Science and Higher Education, contract number MEiN/2023/DIR/3796;  EU Horizon 2020 Teaming grant agreement No 857533; IRAP program of the Foundation for Polish Science; and EU Horizon Europe grant NEARDATA No 101092644.

\printbibliography

\end{document}